\documentclass[traditabstract,letter]{aa}

\usepackage{graphicx}
\usepackage{txfonts}

\begin{document}

\title{A $\sim$4.6~h quasi-periodic oscillation in the BL Lacertae
PKS~2155$-$304?}

\author{P.~Lachowicz\inst{1,2} \and
        A.~C.\ Gupta\inst{3} \and
	H.~Gaur \inst{3} \and
        P.~J.\ Wiita\inst{4,5} }

\institute{   
              Nicolaus Copernicus Astronomical Centre, Polish Academy of
	      Sciences, ul.\ Bartycka 18, 00-716 Warszawa, Poland
	 \and
	      Centre for Wavelets, Approximation and Information
	      Processing, Temasek Laboratories, National University of
	      Singapore, \\ 5A~Engineering~Dr~1, \#09-02 Singapore 117411;
	      pawel@ieee.org
         \and
              Aryabhatta Research Institute of Observational Sciences 	
	      (ARIES) Manora Peak, Nainital, 263129, India; acgupta30@gmail.com
         \and
              School of Natural Sciences, Institute for Advanced Study,
	      Princeton, NJ 08540, USA; wiita@chara.gsu.edu
	 \and
              Department of Physics and Astronomy, Georgia State 			
              University, P.O.\ Box 4106, Atlanta, GA 30302--4106, USA
          }

\date{Received 21.08.2009. Accepted 10.09.2009.}

\abstract{We report a possible detection of an $\sim$4.6-hour quasi-periodic
oscillation (QPO) in the 0.3--10 keV emission of the high-energy peaked blazar
PKS~2155$-$304 from a 64 ks observation by the {\it XMM-Newton} EPIC/pn
detector. We identify a total modulation of $\sim$5\% in the light
curve and confirm that nominal period  by periodogram, structure function and
wavelet analyses. The limited light curve duration allows the capture of only
3.8 cycles of this oscillation and thus precludes a very strong claim for
this QPO, despite a nominally high ($\gtrsim$3$\sigma$)
statistical significance. We briefly discuss models capable of producing an
X-ray QPO of such a period in a blazar.}


\maketitle

\section{Introduction}

Active galactic nuclei (AGN) presumably possess accreting black holes (BHs)
with masses of $10^6$--$10^{10}$ $M_\odot$ and have many similarities to
scaled-up galactic X-ray emitting BH binaries. The presence of quasi-periodic
oscillations (QPOs) is fairly common in both BH
and neutron star binaries in our and nearby galaxies (e.g., Remillard \&
McClintock 2006).  Despite several earlier claims, until recently there were no
convincing cases of QPOs in AGN. A clear detection of a QPO of $\sim$~1 h been
made for the Narrow Line Seyfert 1 galaxy, RE J1034$+$396
(Gierli{\'n}ski et al.\ 2008), and a very strong case for one (also about 1 h)
in a flat spectrum radio quasar, 3C 273, has been reported (Espaillat et al.\
2008), both using data from the {\it XMM-Newton} satellite.

The BL Lacertae object PKS 2155$-$304 ($z = 0.116$) is one of the brightest BL
Lac objects
from UV to TeV energies in the southern hemisphere.  Therefore it is well studied and
known to be rapidly and strongly variable throughout  the  electromagnetic 
spectrum on diverse timescales (e.g., Carini \& Miller 1992; Urry et al.\ 1993;
Brinkmann et al.\ 1994; Marshall et al.\ 2001; Ahronian et al.\ 2005; Dominici,
Abraham \& Galo 2006; Dolcini et al.\ 2007; Piner, Pant \& Edwards 2008; Zhang
2008; Sakamoto et al.\ 2008; and references therein). A giant TeV flare from
this source was observed in July 2006 (e.g., Ahronian et al.\ 2007; Foschini 
et al.\ 2007; Sakamoto et al.\ 2008). Its extreme TeV variability seems to
demand an ultra-relativistic flow (bulk Lorentz factor $\sim$ 50) in at least
portions of the jet that is believed to dominate the emission from this and
other BL Lacs and blazars (Ghisellini \& Tavecchio 2008). This BL Lac
has been the target of several simultaneous multi-wavelength monitoring
campaigns (e.g., Urry et al.\ 1993; Brinkmann et al.\ 1994; Edelson et al.\
1995; Courvoisier et al.\ 1995; Urry et al.\ 1997; Pian et al.\ 1997; Pesce et
al.\ 1997; Foschini et al.\ 2007, 2008). PKS 2155$-$304 is among the few blazars
for which claims of an apparent QPO has been made, with UV and optical
monitoring (from {\it IUE}) over five days possibly having a $\sim$0.7 d
periodicity (Urry et al.\ 1993). Simultaneous X-ray observations tracked those
UV variations fairly well (Brinkmann et al.\ 1994).  

\section{Data and reduction}

We reanalysed archival {\it XMM-Newton} EPIC observations of PKS~2155$-$304
taken on 2006 May 1 
(orbit 1171, ObsID 0158961401). The 0.6--10~keV spectral analysis of this observation has been 
reported in Zhang (2008). We used pipeline products and applied the {\it XMM-Newton} Science 
Analysis System (SAS) version 8.0.0 for the light curve (LC) extraction. We confined our analysis 
to EPIC/pn data as only they were free from soft-proton flaring events and pile-up effects.
In contrast to the data reduction conducted by Zhang (2008), we read out source photons recorded 
in the entire 0.3--10~keV energy band, using a circle of 45 arc-sec
radius centered on the source. Background 
photons were read out from the same size area located about 180 arc-sec off the
source on the 
same chipset.  We finally obtained a source LC (corrected for background flux) of the total
duration, $T\!=\!n\Delta t\!=\!64.1$~ks, of the observation evenly sampled every
$\Delta t\!=\!100$~s.  The mean count rate and rms variability equal 21.9
cts~s$^{-1}$ and 3.1\%, respectively.

\section{Light curve analysis}

A visual inspection of the LC (Figure~\ref{fig:lcaov}a)
suggests that the X-ray emission of PKS~2155$-$304 is modulated by a periodic
component. In order to explore this possibility  we conducted  periodogram, structure
function and wavelet analyses to provide a more complete picture of the observed
variability.

\subsection{Periodogram analyses}

We first used the multi-harmonic AoV periodogram (mhAoV) of Schwarzenberg-Czerny
(1996) to analyze the LC. An extensive description of mhAoV with its
statistical advantages over the classical Lomb-Scargle method and examples of
applications to X-ray time-series analyses can be found in Lachowicz et al.\
(2006). In the periodogram calculations for the signal $x$, we employ Szeg\"{o}
orthogonal trigonometric polynomials as model functions, $x_\parallel$. A series
of $n_\parallel\!=\!2N\!+\!1$ polynomials correspond to the orthogonal
combinations of the $N$ lowest Fourier harmonics where $n_\parallel$ denotes the
number of a model's free parameters. The consistency of the data with the model
is measured by a statistic, $\Theta\equiv\Theta(f)$, that is a function of
frequency. In the mhAoV periodogram $\Theta$ is defined by the Fisher AoV
statistic, namely $\Theta_{\rm AoV}=(n-n_\parallel)
\|x_\parallel\|^2/(n_\parallel\|x-x_\parallel\|^2)$, and follows the
Fisher-Snedecor probability distribution, $F$. Since different periodograms use
different models and statistics, so to facilitate their comparison we convert
$\Theta_{\rm AoV}$ into the false alarm probability, $P_1$. It can be
directly calculated as $P_1=1-F(n_\parallel, n_\perp/n_{\rm corr}; \Theta_{\rm
AoV}/n_{\rm corr})$ where $n_\perp\!=\!n\!-\!n_\parallel$ and $n_{\rm corr}$
defines a number of consecutive observations being correlated (see Sect.~3.1.4
of Lachowicz et al.\ 2006). A direct derivation of $P_1$ returns the
significance of the periodogram peak (centered at the frequency $f_0$) only
under the assumption that any modulation detected has a period,
$P_0\!=\!1/f_0$, which is known in advance. In practice, when a set of
$N_{\rm eff}$
frequencies in a band, $\Delta f$, is scanned for significant periodicities, a
selected peak's significance level should be corrected for multiple trials, so
$P_{N_{\rm eff}} = 1-(1-P_1)^{N_{\rm eff}}$. Since there is no analytical method
that determines $N_{\rm eff}$, we follow Lachowicz et al.\ (2006) and assume a
simple estimate, $N_{\rm eff}\!=\!\min(\Delta f/\delta f, N_{\rm calc}, n)$,
where $\delta f$ defines the peak's FWHM and $N_{\rm calc}$ is the number of
frequencies at which the periodogram is calculated.

\begin{figure}
\includegraphics[angle=0,width=0.475\textwidth]{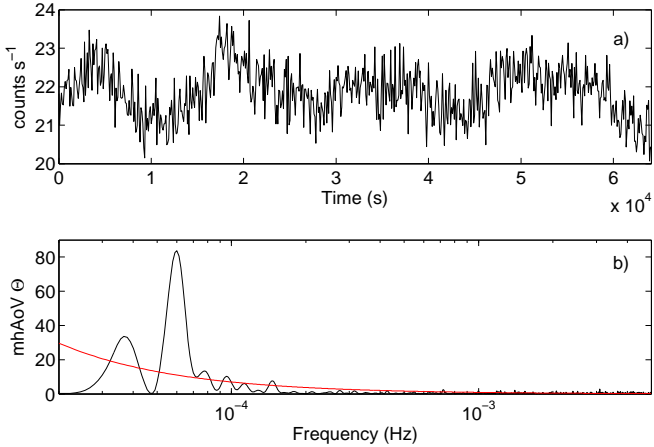}
\caption{(a) The 0.3--10 keV LC of PKS 2155$-$304 as observed
by {\it XMM-Newton} EPIC/pn (orbit 1171) showing a noticeable periodic flux
modulation in the signal; (b) Corresponding mhAoV periodogram revealing a
dominant periodicity at $4.64\pm 0.21$~h. The solid red line denotes the
mean fitted red noise model of a power-law form.}
\label{fig:lcaov}
\end{figure}

\begin{figure}
\includegraphics[angle=0,width=0.475\textwidth]{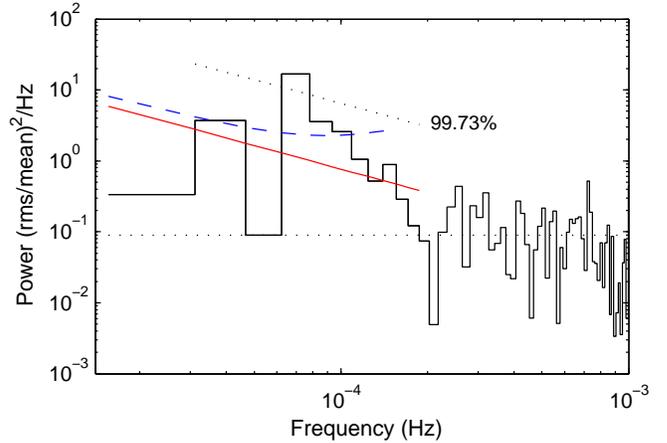}
\caption{Fourier power spectrum of PKS 2155$-$304 (histogram)
and estimated 99.73\% ($3\sigma$) level of confidance (dotted curve).
The solid red line denotes the best fit power-law model of the underlying
red noise with an index of $\alpha=-1.10\pm 0.48$ while the dashed blue line
marks the upper uncertainty in this model. The horizontal dotted line denotes the
expected Poisson noise level.}
\label{fig:psd99}
\end{figure}

For this PKS 2155$-$304 LC we employed the mhAoV assuming a simple model
of $x_\parallel$ with $N\!=\!1$ (a sinusoid). This periodogram analysis
revealed a dominant peak around the frequency of $f\simeq 6\times 10^{-5}$~Hz
(Figure~\ref{fig:lcaov}b). Since the standard frequency
resolution (defined by $1/T$) for frequencies below $f\!<\!10^{-4}$~Hz is
poor, we oversample our periodogram in frequency by a factor of 5. This allows
us to determine the period more precisely and to estimate its $1\sigma$ error
(Schwarzenberg-Czerny 1991). We find $P\!=\!16707\pm 747$~s ($4.64\pm 0.21$~h).
In the error calculation we took into account a correction to the
peak's power from the fitted mean red noise background (assuming a power-law
model) around the suspected QPO frequency (Figure~\ref{fig:lcaov}b; red
line) which reduces $\Theta_{\rm AoV}$ by a factor of about 1.2. We compute
the peak significance and find it to be $P_{N_{\rm eff}}\!<\!10^{-10}$
($\gtrsim\!7\sigma$) using the determined values $\Theta_{\rm AoV}\!=\!70.4$,
$n_{\rm corr}\!=\!1.33$ and $N_{\rm eff}\!=\!n$. Therefore, in terms of the
mhAoV
method, the underlying 4.6~h periodicity stands as statistically significant
even though we capture only 3.8 cycles during the total LC duration. We
find the suspected QPO has a quality factor, $Q\!=\!f/\Delta f\!\simeq\!
12$,  and a fractional rms of 1.6\%.

The underlying continuum that includes red noise can easily give rise to
nominally apparent, but not actually significant, periodicities, particularly
when the dataset only spans a few of the putative periods (Press 1978). 
So in order to more conservatively estimate the significance of this possible
QPO in PKS 2155$-$304 we also employed the approach of Vaughan (2005)
which was used by Gierli{\'n}ski et al.\ (2008) in their analysis of RE
J1034$+$396. This analysis basically involves first measuring the
Fourier periodogram, then estimating the slope of that red noise
continuum contribution through a least-squares fit to the log of the periodogram
and finally estimating the significance of any peaks above that power spectrum
(Vaughan 2005). Although only strictly valid when the spectrum of the underlying
noise is exactly a power-law in frequency, it is a very good approximation as
long as the signal is in the portion of the power spectrum which is close to a
power-law. The results of this analysis are given in Figure~\ref{fig:psd99},
where the peak is just slightly above the 99.73\% confidence level.
This result of $\gtrsim\!3\sigma$ normally would be sufficiently strong to
claim a significant QPO detection. As a comparison, the result using this
approach for the QPO found in the Narrow Line Seyfert 1 galaxy
discussed by Gierli{\'n}ski et al.\ (2008) is $\sim5.6\sigma$ for the better
``Segment 2'' of their LC and $\sim3.4\sigma$ when their entire LC is
considered. Using Monte Carlo simulations we also estimated a
chance probablity of finding a spurious QPO of the $>\!3\sigma$ significance seen in
Fig.\ 2 from that underlying power-law power spectrum to be $4\times 10^{-4}$.

\begin{figure}
\includegraphics[angle=-90,width=0.47\textwidth]{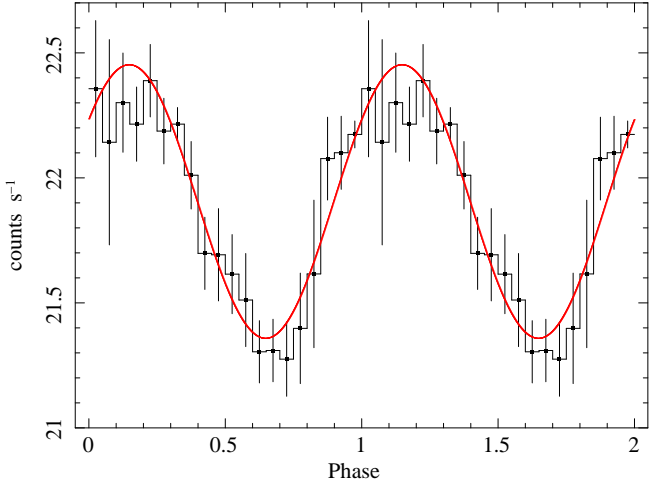}
\caption{Folded 0.3--10 keV LC with period $P=16707$~s
and best fit sinusoid describing flux modulation. Two cycles are displayed
for clarity.}
\label{fig:fold}
\end{figure}

In Figure~\ref{fig:fold} we present a folded LC with a period of
4.64~h re-binned into 20 flux bins per one cycle of the modulation. The best
$\chi^2$ fit of a sinusoidal model to the data is over plotted
and returns a reduced-$\chi^2$ of 0.63 for 17 degrees of freedom. 
We calculate the relative modulation depth, defined by $\phi_{\rm mod}=\!
(y_{\rm max}\!-\!y_{\rm min})/y_{\rm max}$ where $y_{\rm min}$ and $y_{\rm
max}$ denote the minimum and maximum count rates of the fitted model
and find $\phi_{\rm mod}\!=\!4.9\!\pm\!0.3$\%. This is about 40\%
of the periodic modulation at $P\!=\!3733$~s found by
Gierli\'{n}ski et al.\ (2008) in the 0.3--10~keV flux of RE~J1034$+$396.

\begin{figure}
\includegraphics[angle=0,width=0.475\textwidth]{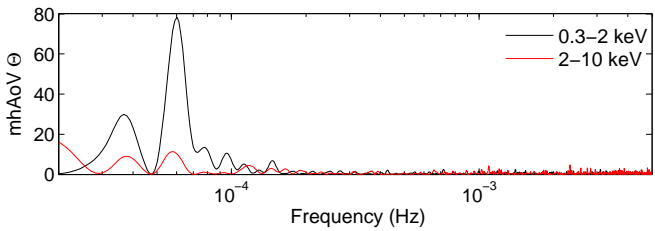}
\caption{Periodograms calculated using mhAoV for PKS 2155$-$304 LCs
corresponding to 0.3--2~keV and 2--10~keV 
emission.}
\label{fig:aov2}
\end{figure}

We also checked how much this detection of $\sim$4.6 h periodicity
depends on energy band. In Figure~\ref{fig:aov2} we compare the mhAoV
periodograms computed for two energy intervals, the ``soft'' 0.3--2~keV and
``hard'' 2--10~keV bands. Surprisingly, while the periodicity is easily
detectable in the soft band, and is still at a high estimated significance
level ($P_{N_{\rm eff}}\!<\!10^{-10}$; mhAoV), it is
visible but not significant in the source's hard X-ray emission.

\begin{figure}
\includegraphics[angle=0,width=0.475\textwidth]{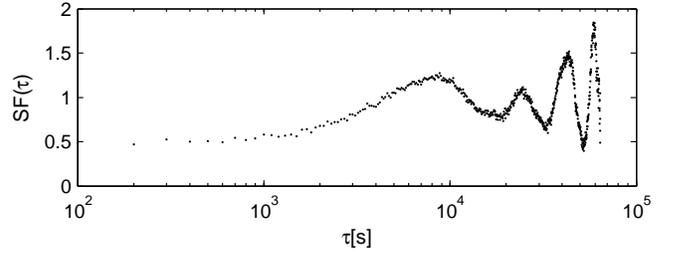}
\caption{Structure Function for the 0.3--10~keV LC of PKS~2155$-$304.
}
\label{fig:sf}
\end{figure}

\subsection{Structure function analysis}

The structure function (SF) is an alternative technique that 
 provides information on the time structure
of a data train and it is able to discern the range of the characteristic time scales that contribute to the fluctuations.
It is less affected by any gaps in the light curves and is free from any constant offset in the time series (e.g., Rutman 1978).
 Introduced by Kolomogorov (1941) the SF
has frequently been applied in astronomy (e.g., Simonetti et al.\ 1985; Paltani et al.\ 1997).  The SF
was recently used by Rani et al.\ (2009) for spotting nearly periodic
fluctuations in the long term X-ray LCs of blazars.

The first-order SF is a time domain technique, defined as ${\rm SF}(\tau) =
\langle [x(t+\tau)-x(t)]^2 \rangle$ where $x(t)\equiv x$ is the LC. For
strictly sinusoidal flux variability with period $P$ the SF has minima for
$\tau$ equal to $P$ and its subharmonics (e.g., Lachowicz et al.\ 2006). Thus,
a SF allows for quick confirmation of the findings provided by the periodogram
techniques. Figure~\ref{fig:sf} displays the SF computed for our X-ray LC of
PKS~2155$-$304. The first and subsequent dips clearly indicate a periodic
component at $\tau\simeq 17000$~s, which is in good agreement with the
periodograms.

\begin{figure}
\includegraphics[angle=0,width=0.475\textwidth]{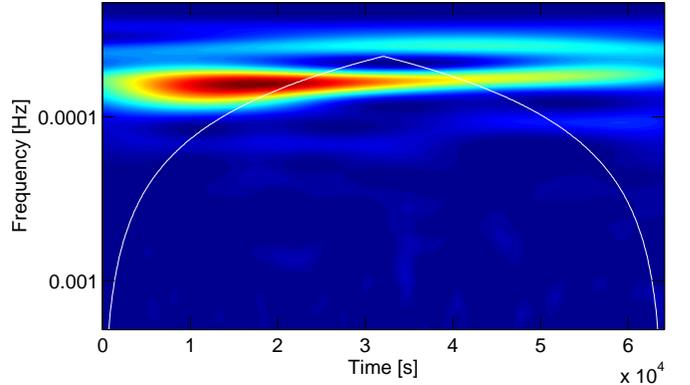}
\caption{Wavelet analysis for the 0.3--10~keV blazar LC, with redder
colors corresponding to stronger signals. The
cone-of-influence is  marked by the solid white line.}
\label{fig:wa}
\end{figure}

\subsection{Wavelet analysis}

We employed the standard wavelet Matlab$^{\mbox{\small
\textregistered}}$ codes of Torrence \& Compo (1998) to scan a time-frequency
plane of the blazar's 0.3--10~keV LC. In the wavelet analysis we used the
Morlet mother function as a basis function. This is a reasonable choice to
capture periodic flux modulations and to provide an idea of their lifetimes and
frequency evolution along the signal. 

Figure~\ref{fig:wa} presents the calculated squared modulus of the continuous
wavelet transform. Two strips of high wavelet power are immediately noticeable
for frequencies $f\!<\!1\!\times\!10^{-4}$~Hz. The frequency of the most
prominent one
agrees very well with our periodogram and structure function indications, i.e.,
at the frequency $f\simeq 6\times 10^{-5}$~Hz of the possible 4.6~h QPO. 

It is worth noting that the wavelet map suggests highest coherence of
that periodicity within the first $\sim$35~ks of the LC which also agrees
with a time-domain based signal analysis. For $t\!>\!35$~ks, the 4.6~h
modulation weakens and its period may stretch out. Within that time interval,
the wavelet map reveals a second flux modulation (i.e. of lower frequency), also
seen in the mhAoV results (Fig.~\ref{fig:lcaov}).  However, much of these
wavelet signals are in the ``cone-of-influence'' where putative modulations are
either too close to the sampling interval or (as in this case) too close to the
total signal length. It is worth noting that this restriction did not apply to
the wavelet analyses employed to support recent claims for X-ray (Espaillat et
al.\ 2008) and optical (Gupta et al.\ 2009) QPOs in other blazars.

\section{Discussion and Conclusions}

A large number of models for the QPOs in Galactic X-ray binaries have
been proposed, and while none  seem to be able to explain all of the
details of the observations, they are all based on fluctuations in, or
oscillations of, accretion disks (e.g., Remillard \& McClintock 2009).
The simplest such models for BHs would attribute the quasi-periods
to particularly strong orbiting hot-spots on the disks at, or close to, the
innermost stable circular orbit allowed by general relativity (e.g., Abramowicz
et al.\ 1991; Mangalam \& Wiita 1993). If such simple models apply in this case,
and the QPO is indeed real, then we would estimate the BH mass for PKS
2155$-$304 to be 3.29 $\times$ 10$^{7} M_{\odot}$ for a non-rotating BH and 2.09
$\times$ 10$^{8} M_{\odot}$ for a maximally rotating BH (e.g., Gupta et al.\
2009; Rani et al.\ 2009). Other possible mechanisms for QPOs in AGNs that could
provide timescales of a few hours can also have a disk origin or can arise from
relativistic jets.  The former class  also includes small epicyclic deviations
in both vertical and radial directions from exact planar motions within a thin
accretion disk (e.g., Abramowicz 2005), oscillations of standing shocks in
transonic flows (e.g., Chakrabarti et al.\ 2004) and trapped pulsational modes
within a disk (e.g., Perez et al.\ 1997; Espaillat et al.\ 2008).  

In the case of PKS 2144$-$304, a disk flux variation is unlikely to
directly produce any detectable QPO because the disk emission is almost
certainly dominated by 
synchrotron emission at low-energy bands and by inverse Compton emission at
higher-energy bands.  Both of these are expected to arise from from
relativistic jets (e.g., Blandford \& K{\"o}nigl 1979; B{\"o}ttcher 2007;
Marscher et al.\ 2008) and their fluxes usually would swamp any disk 
fluxes. Nonetheless, a disk oscillation could trigger a quasi-periodic
injection
of plasma into the jets which could then produce the observed QPO (e.g., Liu et
al.\ 2006).
Some recent work has argued that the nuclear emission in low luminosity 
radio-loud AGNs indicates that the accretion disk is radiatively inefficient
(e.g., Balmaverde et al.\ 2006; Balmaverde \& Capetti 2006).
 Then the emission in even the low luminosity radio-loud AGNs that are likely
the parent population of BL Lacs is dominated by non-thermal emission from the
base of the jet. The observed timescale, $P_{obs}$, of any fluctuation is
likely to be reduced with respect to the rest-frame timescale, $P_{em}$, by the
Doppler factor, $\delta$, and is increased by a factor of (1$+$z). The Doppler
factor depends upon the velocity of the shock propagating down the jet, $V$, and
the angle between the jet and the observer's line-of-sight, $\theta$, as $\delta
= [\Gamma(1 - \beta {\rm cos}\theta)]^{-1}$, where $\beta = V/c$ and $\Gamma =
(1 - \beta^{2})^{-1/2}$. The value of $\delta$ for PKS 2155$-$304 is probably
large, $\sim$30--50  (Urry et al.\ 1997; Ghisellini \& Tavecchio 2008). A shock
propagating down a jet which contains quasi-helical structures, whether in
electron density or magnetic field, can produce a QPO, with successive peaks seen
each time the shock meets another twist of the helix at the angle that provides
the maximum boosting for the observer (e.g., Camenzind \& Krockenberger
1992;
Gopal-Krishna \& Wiita 1992).
Instabilities in jets just might be able to excite such helical modes capable of
yielding fluctuations that are observed to occur on the time scale seen in PKS
2155$-$304 (e.g., Romero 1995). Or they could arise as the jet plasma is launched
in the vicinity of SMBH and thus actually originate in the accretion flow but
become amplified in the jet.  Another very plausible origin for a short-lived
QPO would be turbulence behind the shock in the relativistic jet (e.g. Marscher,
Gear \& Travis 1992), as again intrinsically modest fluctuations could be
Doppler boosted.

While this result for the presence of a QPO in a BL Lac is nominally of reasonably high
statistical significance, the presence of $\!<\!4$ cycles of the putative period in this
dataset means that it can only be considered to be a tantalizing hint, and
not a confirmed case. Detailed discussion of physical mechanisms, along
with results of searches for QPOs in 24 light curves of four other high energy
peaked blazars, will be given in a separate paper (Gaur et al., in preparation).

\acknowledgements
We are deeply grateful to Chris Done for providing the code for the power
spectrum calculation presented in Figure~\ref{fig:psd99}, to Alex 
Schwarzenberg-Czerny for constructive discussion on periodograms and to
Izabela Dyjeci\'{n}ska for graphics advice. This research is based on
observations obtained with XMM-Newton, an ESA science mission with instruments
and contributions directly funded by ESA Member States and NASA.

\end{document}